\documentclass[reqno]{article}

\usepackage{graphicx}
\usepackage{amsmath,amsthm,amssymb}

\chardef\bslash=`\\ % p. 424, TeXbook
\newtheorem{conj}{Conjecture}

\theoremstyle{definition}

\theoremstyle{remark}

\newcommand{\eval}[2][\right]{\relax
  \ifx#1\right\relax \left.\fi#2#1\rvert}

\begin{document}
\title{\bf{Formation of singularities  for equivariant \\
2+1 dimensional wave maps into the two-sphere}}

\author{Piotr Bizo\'n\footnotemark[1]{}{},\;
  Tadeusz Chmaj\footnotemark[2]{}\;,  and Zbis\l aw
  Tabor\footnotemark[3]{}\\{}\\
  \footnotemark[1]{} \small{\textit{Institute of Physics,
   Jagellonian University, Krak\'ow, Poland}}\\
   \footnotemark[2]{} \small{\textit{Institute of Nuclear Physics, Krak\'ow,
    Poland}}\\ \footnotemark[3]{} \small{\textit{Department of Biophysics,
    Collegium Medicum,  Jagellonian University, Krak\'ow,
    Poland}}}
%\date{December 12, 1999 }
%
\maketitle
\begin{abstract}
\noindent In this paper we report on numerical studies of the
Cauchy problem for equivariant wave maps from  $2+1$ dimensional
Minkowski spacetime into the two-sphere. Our results provide
strong evidence for the conjecture that large energy initial data
develop singularities in finite time and that singularity
formation has the universal form of adiabatic shrinking of the
degree-one harmonic map from
 $\mathbb{R}^2$ into $S^2$.
\end{abstract}
\section{Introduction}
 A wave  map is a function
from the Minkowski spacetime $(\mathbb{R}^{n+1},\eta)$ into a
complete Riemannian manifold $(N,g)$,  $U: \mathbb{R}^{n+1}
\rightarrow N$, which is a critical point of the action
\begin{equation}
S(U) = \int g_{AB}\, \partial_a U^A \partial_b U^B  \eta^{ab} \,
d^n\!x\; dt\, .
\end{equation}
The associated Euler-Lagrange equations constitute the system of
semilinear wave equations
\begin{equation}\label{wmap}
\square U^A + \Gamma_{BC}^A(U) \partial_aU^B \partial^a U^C=0,
\end{equation}
where $\Gamma$'s are the Christoffel symbols of the target metric
$g$. Wave maps are interesting both for mathematicians by
providing a simple geometric setting for studying the problems of
global existence and formation of singularities and for physicists
(who call them sigma models) as toy models of extended structures
in field theory (see~\cite{geometric} for a recent review).
In this paper we consider the case where the domain manifold is
 $2+1$ dimensional Minkowski spacetime, $M=\mathbb{R}^{2+1}$, and the target
  manifold
 is
the $2$-sphere,  $N=S^2$, with the standard metric
\begin{equation}
 g = du^2+\sin^2\!u\; d\theta^2.
\end{equation}
 We restrict attention to equivariant maps of the form
 \begin{equation}
u = u(t,r), \quad \theta=\phi,
\end{equation}
where $(r,\phi)$ are the
 polar
coordinates on $\mathbb{R}^2$. The wave map system (\ref{wmap})
reduces then to the semilinear scalar wave equation
\begin{equation}\label{eq}
u_{tt} =u_{rr}+\frac{1}{r}u_r - \frac{\sin(2u)}{2 r^2}.
\end{equation}
The main open question for equation (\ref{eq}) is the issue of
global regularity, namely: do all solutions starting with smooth
initial data
\begin{equation}\label{data}
u(0,r)=u_0(r), \quad u_t(0,r)=u_1(r)
\end{equation}
remain smooth for all times, or do they lose regularity for some
data?
 Our
paper reports on numerical investigations of this problem.

Note that the conserved
 energy associated with solutions of (\ref{eq})
\begin{equation}\label{energy}
E[u]=\pi \int\limits_0^{\infty} (u_t^2+ u_r^2+
  \frac{\sin^2\!{u}}{r^2}) \: r dr
\end{equation}
is invariant under dilations: if
$u_{\lambda}(t,r)=u(t/\lambda,r/\lambda)$, then
$E[u_{\lambda}]=E[u]$. In this sense 2+1 is a critical dimension
for wave maps. Note also that the requirement that energy be
finite imposes a boundary condition at spatial infinity
$u(t,\infty)=k \pi$ ($k=0,1,\dots$) which compactifies
$\mathbb{R}^2$ into $S^2$ and thus breaks the Cauchy problem into
infinitely many disconnected topological sectors labeled by the
degree $k$ of the map $S^2\rightarrow S^2$.

 Let us recall what is known
rigorously about the problem. Besides the routine local existence
proof, the only global result is that there is a unique smooth
solution for all times provided that the initial energy is
sufficiently small~\cite{geometric}. In order
 to obtain global
existence without the assumption of small energy it would be
sufficient
 to show that the energy cannot concentrate at the hypothetical
singularity.  This kind of result was derived by Shatah and
Tahvildar-Zadeh~\cite{shatah1} for convex targets and then
extended by Grillakis~\cite{gril} for  nonconvex targets with
arbitrarily narrow neck. However, the Morawetz identity, which is
the basic tool in proving nonconcentration of energy, loses the
desired positivity properties for the geometry of two-sphere. This
raises the question: is the restriction on admissible targets
assumed in~\cite{gril} only of technical nature or is it
essential? In particular, is concentration of energy  possible in
the case of two-sphere as target?

We present numerical evidence  that for large energy solutions
 the energy does in fact concentrate
and consequently the solutions blow up in finite time. We show
that the process of energy concentration proceeds via adiabatic
evolution along the one-parameter family of dilations of the
degree-one static solution. In this sense the shape of blowup is
universal. The rate of blowup, determined by the asymptotic speed
of adiabatic evolution, is slower than that predicted by the
geodesic approximation and goes to zero as the singularity is
approached.

The  paper is organized as follows. In section~2 we derive static
solutions  and study their stability. As mentioned above these
solutions play an essential role in the process of singularity
formation. For completeness, in section~3 we discuss singular
 self-similar solutions and argue that they play no role in the
 Cauchy problem. The main body of the paper is contained in section~4
 where we present the results of
 numerical investigations. On the basis of these results we
 formulate
  three conjectures about  the nature of singularity formation
   in the model. Finally, in section~5 we comment on some earlier work
   on this problem and point out some open problems.
\section{Static solutions}
 Static solutions of equation (\ref{eq})
 can be interpreted as spherically symmetric
harmonic maps from the Euclidean space $\mathbb{R}^2$ into $S^2$.
They satisfy the ordinary differential equation
\begin{equation}\label{static}
u''+\frac{1}{r} u' - \frac{\sin(2 u)}{2 r^2} =0,
\end{equation}
where $'=d/dr$. The obvious constant solutions of (\ref{static})
are $u=0$ and $u=\pi$; geometrically these are maps into the north
and the south pole of $S^2$, respectively. The energy of these
maps attains the global minimum $E=0$. Another constant solution
is the equator map $u=\pi/2$ but this solution is singular and has
infinite energy. The fact that equation (\ref{static}) is scale
invariant does not exclude nontrivial regular solutions with
finite energy (Derrick's argument is not applicable) and, in fact,
such solutions are well-known both in the mathematical literature
as harmonic maps from $\mathbb{R}^2$ into $S^2$ and in the physics
literature as instantons in the two-dimensional euclidean sigma
model. They can be derived in many ways. One possibility is to use
a Bogomol'nyi-type argument which goes as follows. Let $x=\ln{r}$
and $U(x)=u(r)$. Then, assuming staticity
\begin{equation}\label{bogom}
E[u]=\pi\int\limits_0^{\infty} ({u'}^2+
  \frac{\sin^2\!{u}}{r^2}) \: r dr =\pi \int\limits_{-\infty}^{\infty}
  ({U'}^2 + \sin^2\!{U}) \:  dx = \pi \int\limits_{-\infty}^{\infty}
  (U'- \sin\!{U})^2 \: dx - 2\pi \cos{U}
   \Bigr\rvert^{\infty}_{-\infty}
\end{equation}
Thus, in the topological sector $k=1$ the energy attains the
minimum, $E=4\pi$, on the solution of the first order equation $
U'-\sin\!{U}=0$, which is $U(x)=2 \arctan(e^{x})$ up to
translations in $x$.  Therefore
\begin{equation}\label{ssol}
u_S(r)=2 \arctan(r)
\end{equation}
is the static degree-one solution (the problem has reflection
symmetry so, of course, $-u_S(r)$ is also the solution). By
dilation symmetry, the solution $u_S(r)$
 generates the orbit
of static solutions  $u_S^{\lambda}(r)=u_S(r/\lambda)$. We remark
in passing that  the solution (\ref{ssol}) can be alternatively
obtained in the elegant geometric way by taking the identity map
between two-spheres and making the stereographic projection.

  We consider now the  linear stability of the static solution $u_S(r)$.
   Inserting
  $u(t,r)=u_S(r)+e^{ikt} v(r)$ into (\ref{eq}) and linearizing, we get the
  eigenvalue problem (the radial Schr\"odinger equation)
\begin{equation}\label{pertstat}
  L v = \left(-\frac{d^2}{dr^2}-\frac{1}{r}\frac{d}{dr} +  V(r)\right) v = k^2 v,
\end{equation}
where
\begin{equation}\label{potential}
V(r)=\frac{\cos(2 u_S)}{r^2}=\frac{1- 6 r^2 + r^4}{(1+r^2)^2 r^2}.
\end{equation}
This potential  has no bound states as can be shown by the
following standard argument.
 Consider the perturbation induced by dilation
\begin{equation}\label{trick}
v_0(r)=-\frac{d}{d \lambda} u_S^{\lambda}(r)
\Bigr\rvert_{\lambda=1} = r u_S'(r) = \frac{2 r}{1+r^2}.
\end{equation}
This is the solution to $k^2=0$ (so called zero mode). The fact
that the zero mode $v_0(r)$ has no nodes implies by the standard
result from Sturm-Liouville theory that there are no negative
eigenvalues, and \emph{eo ipso} no unstable modes around $u_S(r)$.
Note that the zero mode is not a genuine eigenfunction because it
is not square integrable. Therefore the  operator $L$ has the
purely continuous spectrum $k^2 \geq 0$.
\section{Nonexistence of self-similar solutions} Since equation
(\ref{eq}) is scale invariant, it is  natural
  to look for self-similar
solutions of the form
\begin{equation}\label{ansatz}
u(t,r)=f\left(\frac{r}{T-t}\right), \end{equation}
 where $T$ is a
positive constant.  Substituting this ansatz  into (\ref{eq}) one
obtains the ordinary differential equation
\begin{equation}\label{ss}
\frac{d^2
f}{d\rho^2}+\left(\frac{1}{\rho}-\frac{\rho}{1-\rho^2}\right)
\frac{df}{d\rho} -\frac{\sin(2f)}{2 \rho^2 (1-\rho^2)} = 0.
\end{equation} Let us consider equation (\ref{ss})
inside the past light cone of the point $(t=T,r=0)$, that is for
$\rho \in [0,1]$. It is well-known that there are no solutions
which are analytic at the both ends of this
interval~\cite{geometric}. However, it is not-well-known that
there are solutions which are less regular. They can be easily
derived by setting $\rho=1/\cosh(y)$, so that (\ref{ss})
simplifies to
\begin{equation}\label{xeq}
\frac{d^2 f}{dy^2}-\frac{1}{2} \sin(2 f) =0.
\end{equation}
This equation is solved (up to translations in $y$) by $f(y)=2
\arctan(e^y)$, so in terms of $\rho$ we get a one-parameter family
of self-similar solutions
\begin{equation}\label{sss}
f_{\alpha}(\rho)= 2 \arctan\left(\frac{\alpha
\rho}{1+\sqrt{1-\rho^2}}\right).
\end{equation}
These solutions are analytic at $\rho=0$ but they are not
differentiable at $\rho=1$ (and consequently have infinite energy).
 Since such  solutions cannot develop
from smooth initial data inside the whole past light cone of the
singularity, they are not expected to play any role in the Cauchy
problem. The numerical results described below support this
expectation. In this respect the wave maps in 2+1 dimensions are
completely different from the wave maps in 3+1 dimensions where a
stable analytic self-similar solution determines the process of
singularity formation~\cite{my}.
\section{Numerical results}
 We have solved numerically the Cauchy
problem (\ref{eq}), (\ref{data}) for various one-parameter
families of initial data which interpolate between small and large
energy. The details of our numerical methods are described in the
appendix. The results described below are universal in the sense
that they do not depend on the choice of the family of initial
data (nor on the topological sector). For concreteness we present
them for the degree-zero initial data of the form
\begin{equation}\label{gaus}
u(0,r)=A \left(\frac{r}{R}\right)^3
\exp\left[-\left(\frac{r-R}{\delta}\right)^4\right], \quad
u_t(0,r)=0,
\end{equation}
where the amplitude $A$, the radius $R$, and the width $\delta$
are free parameters. Below we fix $R=2$, $\delta=0.4$ and vary
$A$. We emphasize that the amplitude is by no means distinguished:
any parameter which controls the energy of the initial data could
be varied.  Note that regularity at the center requires that
$u(t,0)=0$ for all $t<T$, where $T$ is a time when the first
singularity (if there is any) develops at $r=0$. Since the initial
momentum is zero, the initial profile splits into two waves,
ingoing and outgoing, traveling with approximately unit speed. The
evolution of the outgoing wave has nothing to do with singularity
formation so we shall ignore it in what follows. The behavior of
the ingoing wave depends on the amplitude $A$. For small
amplitudes, the ingoing wave shrinks, reaches a minimal radius,
and then expands to infinity leaving behind the zero energy region
(see figure~1). When $A$ increases, the minimal radius at which
the wave bounces back decreases and seems to go to zero for some
critical value of the amplitude $A^{\star}$. Finally, for the
supercritical amplitudes $A>A^{\star}$, the wave does not bounce
back and keeps shrinking to zero size in finite time. More
precisely, we observe that evolution of the wave near the center
(so called inner solution) is well approximated by  the degree-one
static solution $u_S$ with a time-dependent scale factor $\lambda$
\begin{equation}\label{adiabatic}
  u(t,r) \approx -2 \arctan\left(\frac{r}{\lambda(t)}\right).
\end{equation}
We shall refer to this formula as the adiabatic approximation.
Using the adiabatic approximation the evolution of the ingoing
wave can be described as follows.  For subcritical amplitudes
$A<A^{\star}$ the scale factor $\lambda(t)$ decreases, attains a
minimum $\lambda_{min}$, and then increases. When $A \rightarrow
A^{\star}$, then $\lambda_{min} \rightarrow 0$. For supercritical
amplitudes $A>A^{\star}$, the scale factor decreases monotonically
to zero in finite time. As follows from (\ref{adiabatic}),
$u_r(t,0) \sim \lambda^{-1}(t)$, hence for the supercritical
solutions the gradient blows up at the center in finite time.
Various aspects of this behavior and some numerical details are
shown in figures~2, 3, 4, and 5.

We have not been able to develop a rigorous mathematical
understanding of the adiabatic approximation (\ref{adiabatic}) but
we can make some heuristic arguments which help understand the
observed behavior. Let us define a similarity variable
$\eta=r/\lambda(t)$. Substituting $u=u(t,\eta)$ into equation
(\ref{eq}) gives
\begin{equation}\label{eta}
- \ddot u + (1- \dot \lambda^2 \eta^2) u'' +\left[1+ (\lambda
\ddot \lambda -2\dot \lambda^2) \eta^2\right] \frac{u'}{\eta}
-\frac{\sin(2 u)}{2 \eta^2} =0,
\end{equation}
where $\dot{}=\partial/\partial t$, $'=\partial/\partial\eta$. In
order to "explain" the observed behavior we make two assumptions.
The first assumption, which is the essence of adiabaticity, says
that the dynamics of the solution near the center is slaved in the
varying scale $\lambda(t)$. This implies that we can neglect the
term $\ddot u$ in (\ref{eta}) which refers to the \emph{explicit}
time dependence. The second assumption concerns the rate of blowup
and says that
\begin{equation}\label{rate}
  \frac{\lambda(t)}{T-t} \rightarrow 0 \quad \mbox{as} \quad t
  \nearrow T.
\end{equation}
Eq.(\ref{rate}) implies that the terms involving the time
derivatives of $\lambda$ in Eq.(\ref{eta}) tend to zero as $t
\nearrow T$. After dropping these terms and the $\ddot u$ term,
Eq.(\ref{eta}) becomes formally the same as Eq.(\ref{static}) and
therefore is solved by $u_S(\eta)$. Although this \emph{ad hoc}
explanation is certainly not satisfactory, it is fully consistent
with the numerics. In particular, it explains  why the "amount" of
the inner solution which is approximated by (\ref{adiabatic})
increases as the wave shrinks (see figure~5).

Of course, it would be very interesting to find an exact rate of
blowup. We obtain a reasonable fit to  the power law behavior
\begin{equation}\label{powerlaw}
  \lambda(t) \sim (T-t)^{\alpha} \quad \mbox{as} \quad t
  \nearrow T,
\end{equation}
with the anomalous exponent $\alpha \approx 1.1\pm 0.05$ (see figure~6).
However, in view of the limited resolution of our numerics near
the blowup and the lack of theoretical arguments behind
(\ref{powerlaw}) we caution the reader not to take
(\ref{powerlaw}) as a serious prediction; in particular we cannot
rule out logarithmic corrections to the power law behavior.

 Now, on the
basis of the numerical studies just described we would like to put
forward three conjectures which summarize the main points of our
findings.
\begin{conj}[On blowup for large data] For initial data (\ref{data}) with
 sufficiently large energy,
the solutions of equation (\ref{eq}) blow up in finite time  in
the sense that the derivative $u_r(t,0)$ diverges as $t \nearrow
T$ for some $T>0$.
\end{conj}
\begin{conj}[On blowup profile]
 Suppose that the solution $u(t,r)$ of the initial value problem
 (\ref{eq}),(\ref{data}) blows up at
some time $T>0$. Then, there exists a positive function
$\lambda(t)\searrow 0$ for $t\nearrow T$ such that
\begin{equation}\label{limit}
  \lim_{t\nearrow T} u(t,\lambda(t) r) = \pm u_S(r) \quad for \quad r>0.
\end{equation}
\end{conj}
\begin{conj}[On energy concentration]
 Suppose that the solution $u(t,r)$ of the initial value problem
 (\ref{eq}),(\ref{data}) blows up at
some time $T>0$. Define the kinetic and the potential energies at
time $t<T$ inside the past light cone of the singularity by
\begin{equation}\label{potential}
E_K(t)=\pi \int\limits_0^{T-t} u_t^2\: r dr \quad \mbox{and} \quad
 E_P(t)=\pi \int\limits_0^{T-t} (u_r^2+
  \frac{\sin^2\!{u}}{r^2}) \: r dr.
\end{equation}
Then:
\begin{description}
\item[(i)] the kinetic energy tends to zero at the singularity
\begin{equation}
\label{conj3}
  \lim_{t \nearrow T} E_K(t) = 0;
\end{equation}
\item[(ii)] the potential energy equal to the energy of the static
solution $u_S$ concentrates at the singularity
\begin{equation}
\label{conj4}
  \lim_{t \nearrow T} E_P(t) = E[u_S]=4 \pi.
\end{equation}
\end{description}
\end{conj}
\noindent We have already discussed the evidence for Conjectures~1
and 2. Conjecture~3 is basically the consequence of
(\ref{adiabatic}) and (\ref{rate}). To see this let us substitute
$u_S(r/\lambda(t))$ into (\ref{potential}) to get
\begin{equation}\label{potential2}
  E_K(t)=\pi \dot\lambda^2 \int\limits_0^{\frac{T-t}{\lambda(t)}}
{u_S'}^2 \: r^3 dr \quad \mbox{and} \quad E_P(t)=\pi
\int\limits_0^{\frac{T-t}{\lambda(t)}} ({u_S'}^2+
  \frac{\sin^2\!{u_S}}{r^2}) \: r dr.
\end{equation}
Assuming (\ref{rate}), the upper limits in these integrals tend to
infinity as $t\nearrow T$, so (\ref{conj3}) and (\ref{conj4})
follow (notice that the integral in $E_K$ diverges
logarithmically). Conjecture~3 means that as the blowup is
approached the excess energy above the energy of the static solution
flows outward from the inner region. This is clearly seen in our
simulations.

 We remark that
 an averaged weak version of (\ref{conj3})
\begin{equation}
 \lim_{\epsilon \rightarrow 0} \frac{1}{\epsilon} \int\limits_{T-\epsilon}^T E_K(t) dt = 0
\end{equation}
was proved  by Shatah and Tahviladar-Zadeh~\cite{shatah2}.
\section{Final remarks}
We would like to comment on two papers  by Piette and
Zakrzewski~\cite{wojtek} and by Linhart~\cite{lin}  which were
devoted to the adiabatic evolution  in the degree-one topological
sector. These authors considered initial data which have the shape
of the static solution and a nonzero momentum directed inwards.
They observed adiabatic shrinking (\ref{adiabatic}) with the scale
factor $\lambda$ changing approximately linearly in time. This was
"explained" by the geodesic approximation (an old idea due to
Manton~\cite{nick}) as follows.
 Substituting the ansatz (\ref{adiabatic})
into the action one obtains the effective action for the scale
factor $\lambda(t)$. The potential energy part does not depend on
$\lambda$, so only the kinetic energy part contributes to the
effective action,
\begin{equation}\label{effec}
S_{eff}[\lambda] =  \int dt \int\limits_0^{\infty} u_t^2\;   r dr
\sim \int dt \; \dot\lambda^2 \int\limits_0^{\infty} \frac{r^3
dr}{(\lambda^2+r^2)^2} \sim c\int \dot\lambda^2 dt,
\end{equation}
where the "constant" $c$ is logarithmically divergent\footnote{In
the language of geodesic approximation this divergence means that
the volume of the moduli space is infinite which nota bene  is
equivalent to the fact that the zero mode is not square
integrable. In the literature one can find  statements that  zero
modes which are not square integrable are "frozen" by infinite
inertia.
  The papers~\cite{wojtek} and~\cite{lin} demonstrate that
these statements, based on the naive picture of (\ref{effec}) as
the action for the free particle with infinite mass, are wrong.}.
As long as the geodesic approximation is used to model blowup,
this divergence is irrelevant because it can be removed by
truncating the action at some large radius. By finite speed of
propagation, such a truncation cannot affect the blowup. Thus, the
effective action (\ref{effec}) yields the scale factor going to
zero linearly $\lambda \sim T-t$ as $t\nearrow T$. To verify the
accuracy of this approximation, Piette and Zakrzewski have solved
the Cauchy problem numerically and got $\lambda \sim
(T-t)^{1+\epsilon}$, where $\epsilon$ is a positive number of the
order of $0.1$. The smallness of $\epsilon$ was interpreted in
favor of the geodesic approximation. Although we have confirmed
these results numerically (see the discussion above and figure~6),
we disagree with the authors of ~\cite{wojtek} and~\cite{lin}
regarding the accuracy of the geodesic approximation. As we wrote
above, the assumption (\ref{rate}) is crucial for the validity of
the adiabatic approximation. In contrast, the linear decay of
$\lambda$, predicted by the geodesic approximation, is
inconsistent with the observed adiabatic evolution, and, as
follows from (\ref{potential2}), gives the wrong prediction on
energy concentration at the blowup which contradicts Conjecture~3
and
  theorem (28). The reason for which
the geodesic approximation fails to capture these crucial features
of blowup is easy to understand: this approximation completely
neglects radiation which is essential in expelling the excess
energy from the inner region.

Besides the obvious problem of proving Conjectures 1, 2, and 3,
the research presented here raises a number of questions. Most
interesting among them, in our opinion, are:\\ \noindent 1. What
mechanism selects the time evolution of the scale factor
$\lambda$? In answering this question the methods of center
manifold theory might be useful. In particular, using weakly
nonlinear stability analysis it should be possible to derive the
amplitude equations for the nonlinear evolution of the zero mode.
The problem shares many
 features with the problem of blowup for the nonlinear Schr\"odinger equation
 in two spatial dimension. It is feasible that the techniques of asymptotic
 matching used there~\cite{papa} could be also applied to our
 problem.\\
2. What is the evolution at the threshold for singularity
formation? What does the fine-tuning accomplish dynamically? The
model does not fit into the framework of "standard"
threshold behavior where a codimension-one stable manifold of a
certain critical solution separates blowup from
dispersion~\cite{icmp}. Unless
 more accurate fine-tuning 
would reveal a new universal behavior very close to the threshold
(which is unlikely), it seems that solutions evolving from exactly critical
initial data also blow up in the adiabatic manner but at a much slower rate
(see Fig.~6).
 \\
3.
 To what extent
the results are specific to the equivariance ansatz; is the blowup
stable under general perturbations?

We hope to be able to say more about these problems in
future publications.
\section*{Acknowledgments}
PB thanks Shadi Tahvildar-Zadeh for helpful remarks. We also thank
Jim Isenberg and Steve Liebling for discussions and for informing
us that their independent studies of the problem are in agreement
with our results.
 When this
paper was almost finished, we received the preprint "Equivariant
wave maps in two space dimensions" by Michael Struwe in which a
version of Conjecture~2 is proved. We are grateful to Michael
Struwe for letting us know about his result. This research was
supported in part by the KBN grant 2 P03B 010 16.
\section*{Appendix: Numerical methods}
\newcounter{zahler}
\renewcommand{\thesection}{\Alph{zahler}}
\renewcommand{\theequation}{\Alph{zahler}.\arabic{equation}}
\setcounter{zahler}{1}%{2}
\setcounter{equation}{0}
In order to solve equation (\ref{eq}) numerically we rewrite it as
the first order system in time:
\begin{subequations}
\begin{eqnarray}
u_t &=& v,\\
 v_t &=& u_{rr} + \frac{1}{r} u_r - \frac{\sin(2 u)}{2 r^2}.
\end{eqnarray}
\end{subequations}
We solve this system by finite differencing. To ensure regularity
at the origin we require that $u(r,t) = O(r)$ for $r \rightarrow
0$, from which the inner boundary conditions follow: $u(0,t)=0,
v(0,t)=0$. As the outer boundary condition we impose an
approximate outgoing wave condition. A naive centered difference
scheme applied to the right hand side of (A.1b) would trigger an
instability near $r=0$. To avoid this, we use a scheme which is
natural for the operator
\begin{equation}
 {\cal L} u = u_{rr}
+\frac{1}{r} u_r = \frac{1}{r} \partial_r ( r
\partial_r u),
\end{equation}
 and takes the form
\begin{equation}
  {\cal L} u \approx \frac{1}{r}
           \left[\frac{1}{h} \left( (r+\frac{h}{2}) 
	   \frac{u(r+h) - u(r)}{h} -
                   (r-\frac{h}{2}) \frac{u(r) - u(r-h)}{h}
                   \right) \right],
                   \end{equation}
 where $h$ denotes the spatial mesh size.

For time evolution we use a standard leapfrog scheme. In that way
we obtain a scheme which is second order accurate in space and
time.

Preliminary results obtained on a uniform, fixed grid show that
the most interesting inner solution is well approximated by the
static solution $u_S$ with a time-dependent scale factor $\lambda$
(see equation (19)). Therefore, to follow this solution we have to
change the scale of spatial resolution in time and keep it roughly
proportional to $\lambda$, at least as long as $\lambda$
decreases. To this end we apply an adaptive  algorithm in which
both the mesh size and the time step are refined as the solution
shrinks. We start with a uniform grid covering an interval $(0,R)$
with some initial resolution characterized by a mesh size
$h=\Delta r_0$. We continue the evolution of the system on this
grid with a time step $\Delta t_0$ as long as $ u_r(0,t) h \le C $
where $C$ is some small fixed constant (spatial tolerance factor).
When this inequality is violated we refine the original grid on
the interval $(0,R/2)$ by  covering it with the resolution $\Delta
r_1 = \Delta r_0/2$. The values of functions at the new points,
not defined on the parent grid, are obtained by interpolation.
From that time on we continue the evolution on the finer grid with
the time step $\Delta t_1 = \Delta t_0/2$. Iterating this process
several times we get the resolution adaptively adjusted to the
solution.

In order to make sure that the numerical results are reliable, we
have reproduced them using a different implicit
finite-differencing scheme in which we used $\ln(r)$ as the
spatial variable.
\newpage
\pagestyle{empty}
\begin{figure}
 \centering
\includegraphics[width=\textwidth]{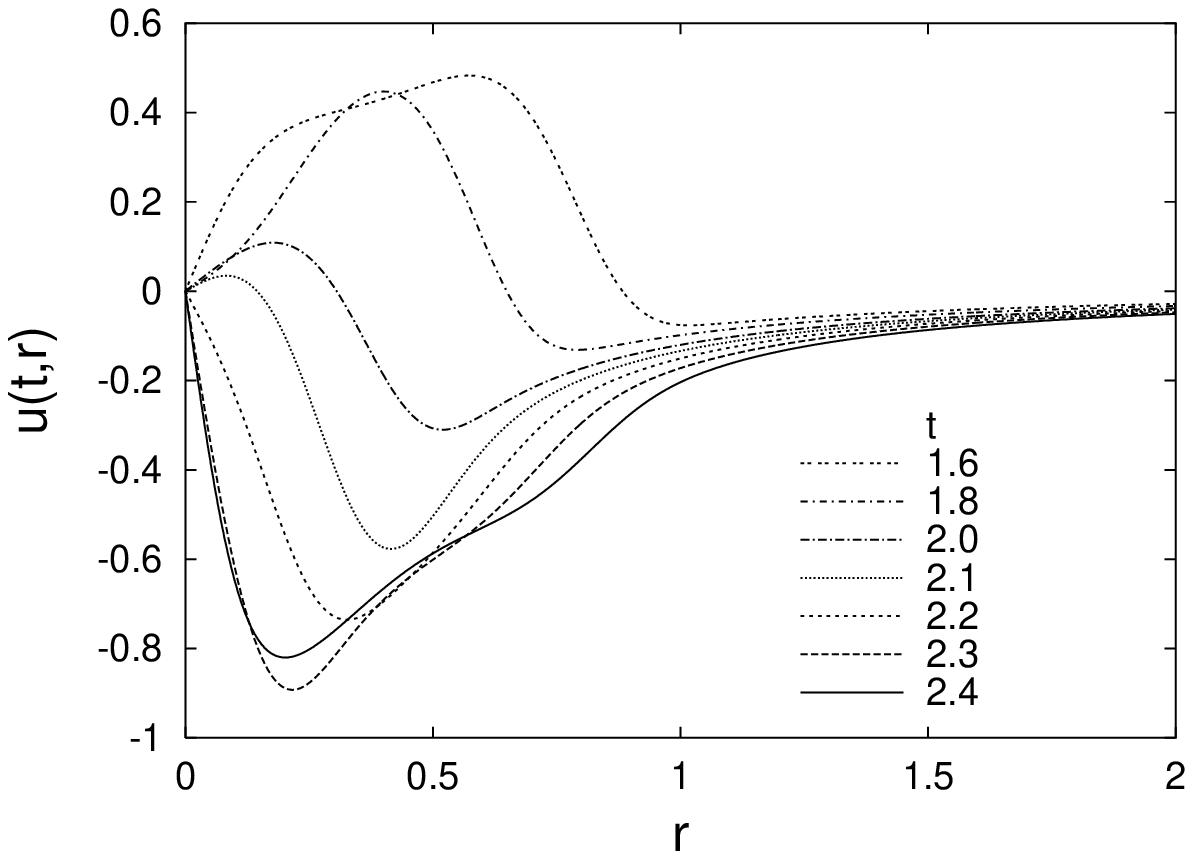}\\
\vspace{0.1in}%
\centering
\includegraphics[width=\textwidth]{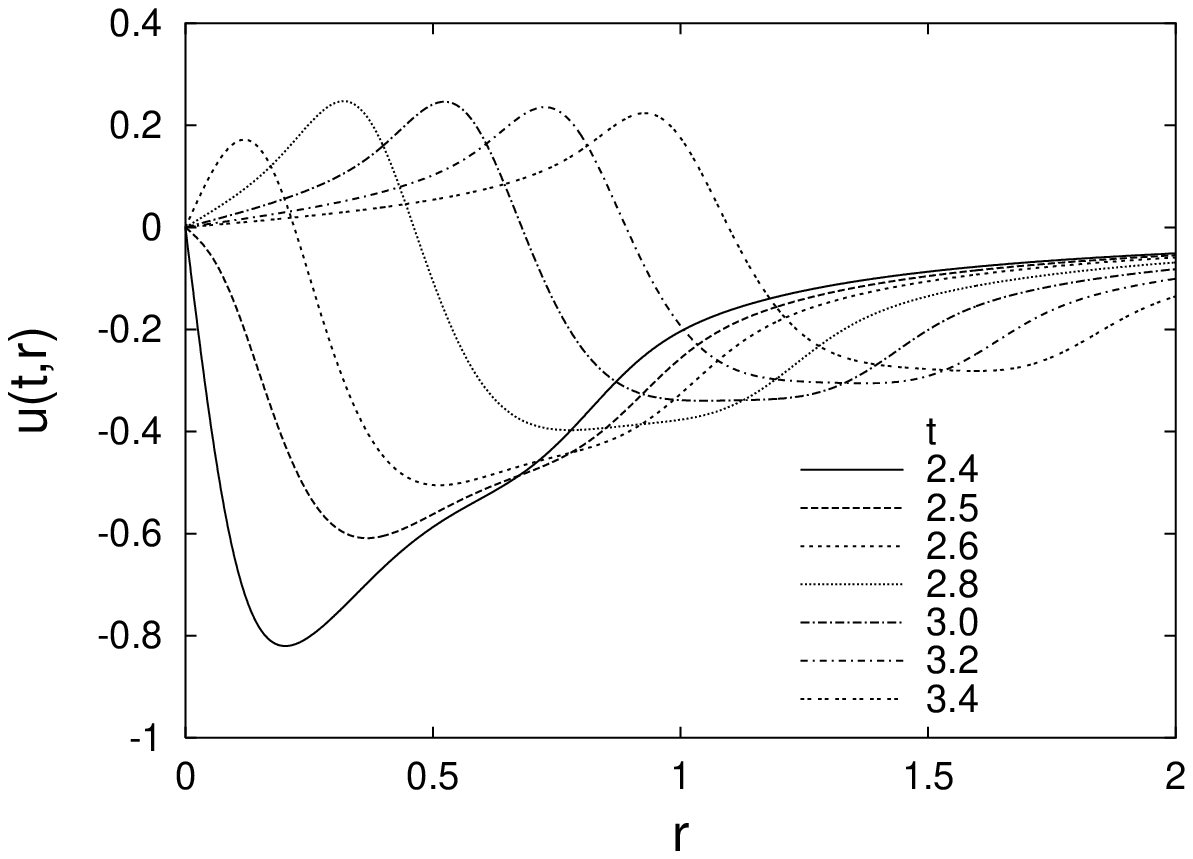}
\caption{Snapshots of the evolution of initial data (\ref{data})
with small amplitude $A=0.5$. The ingoing wave bounces back and
disperses. The minimal radius is attained at $t \approx 2.4$.}
\end{figure}
\begin{figure}
 \centering
\includegraphics[width=\textwidth]{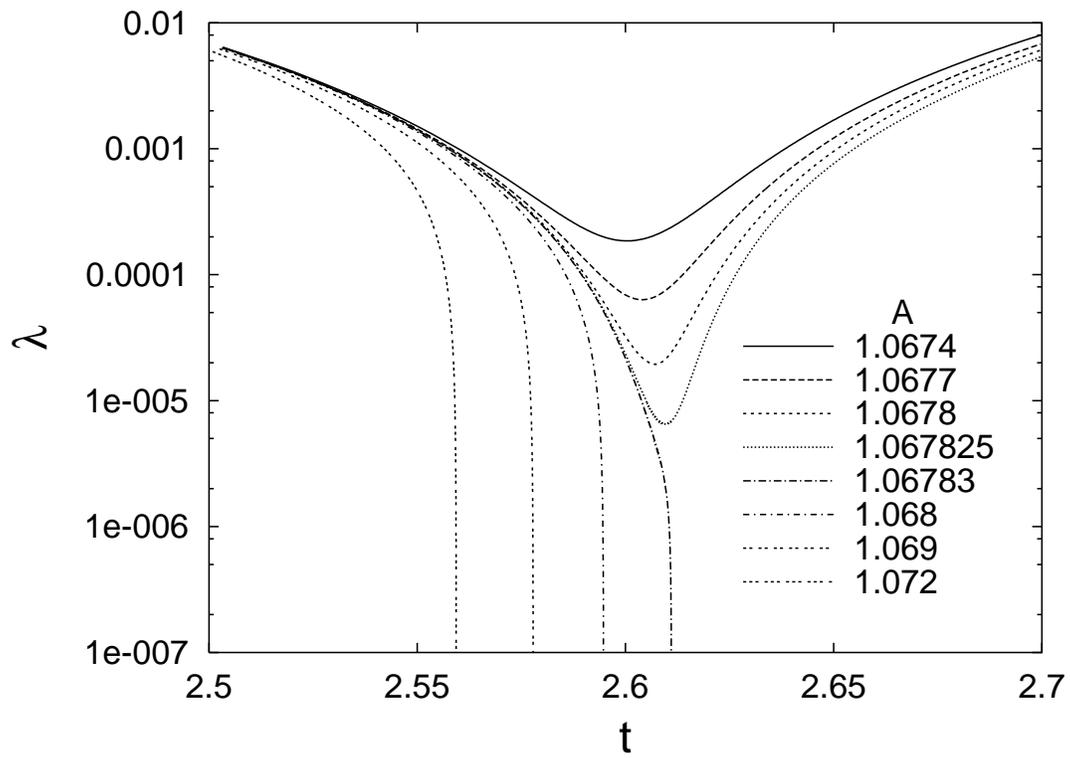}
\caption{The scale factor $\lambda(t)$ for a sequence of solutions
with nearly critical initial amplitudes.  Numerically,
$\lambda(t)$ was calculated from the formula
$u_r(t,0)=-2/\lambda(t)$.  The critical amplitude was estimated to
be $A^{\star} \approx 1.0678281$.}
\end{figure}
\begin{figure}
 \centering
\includegraphics[width=0.9\textwidth]{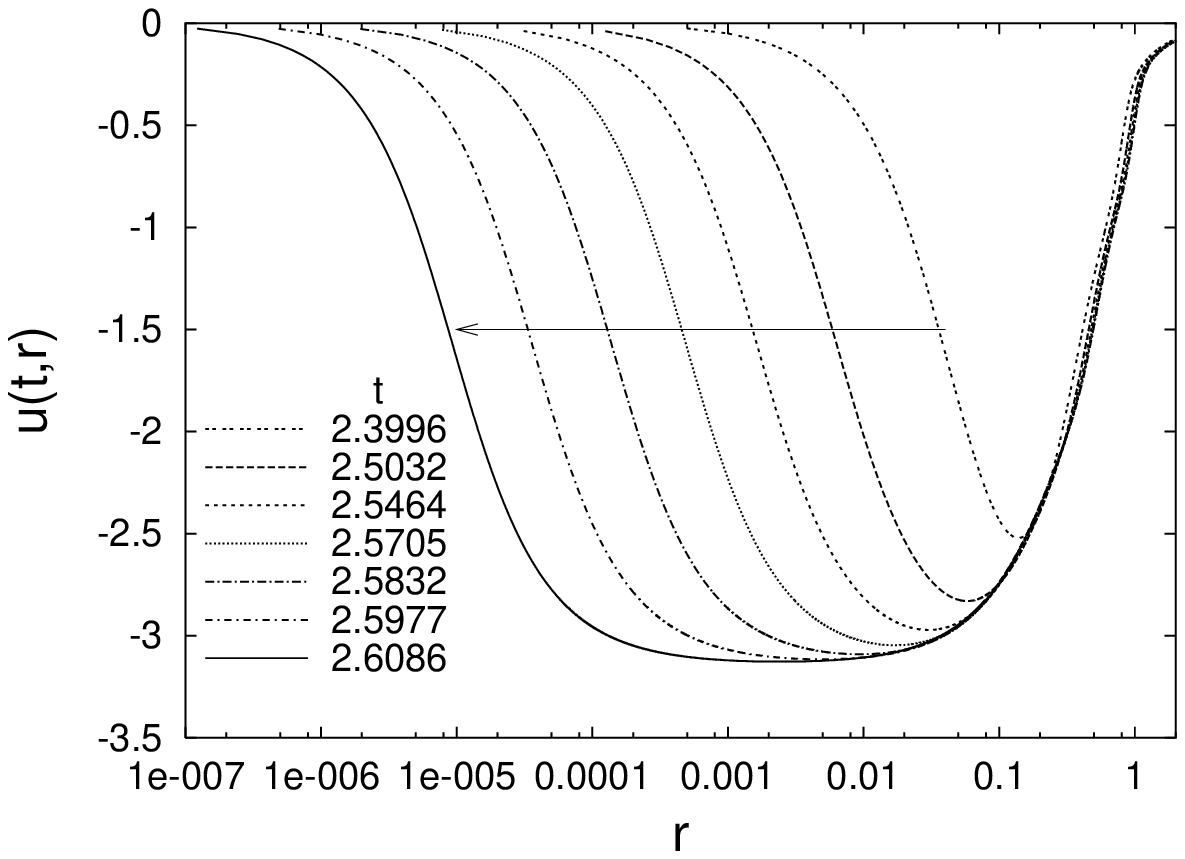}\\
\vspace{0.1in}%
\centering
\includegraphics[width=0.9\textwidth]{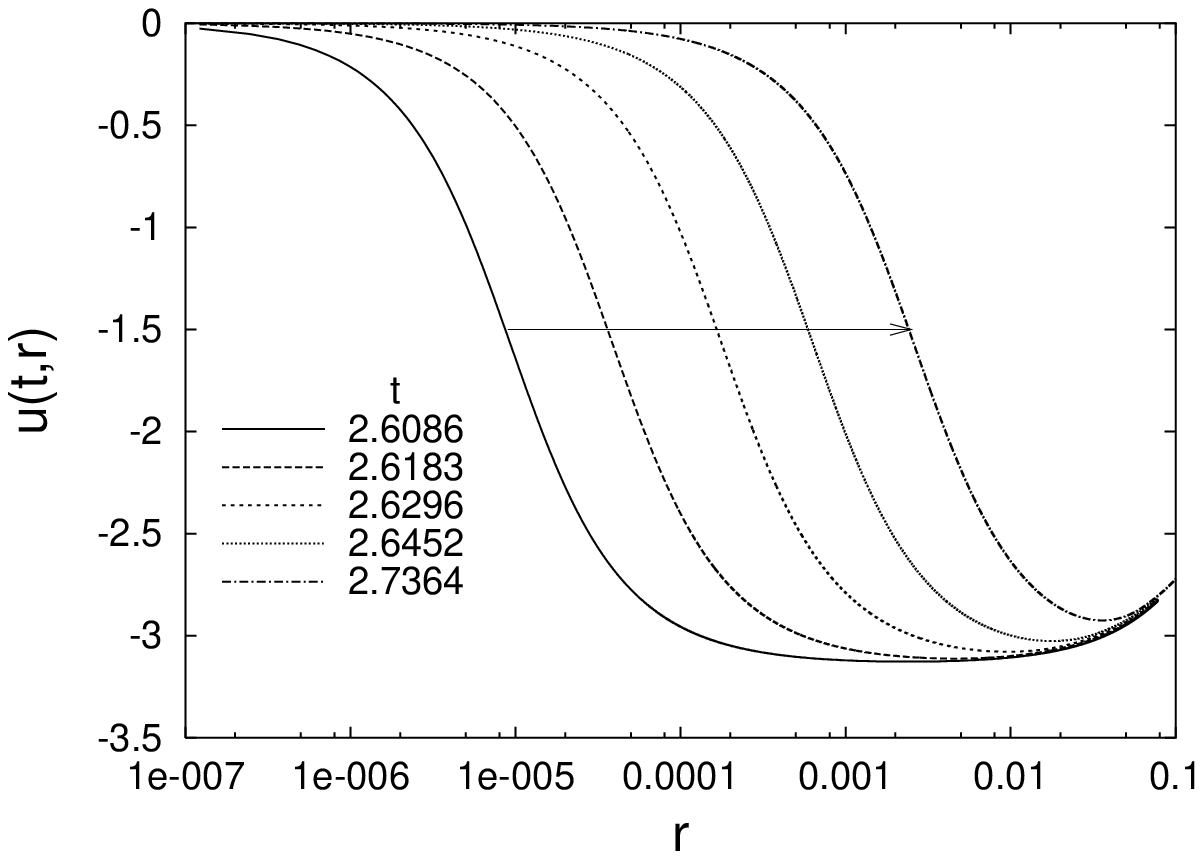}
\caption{Snapshots of the evolution for  the marginally
subcritical amplitude $A=1.06782$. The upper plot shows the
shrinking phase while the lower plot shows the expanding phase.
The horizontal arrows indicate the direction of motion. For
sufficiently small $r$ all profiles have the shape of the suitably
rescaled static solution. The scale factor attains the minimum
$\lambda_{min} \approx 0.9308\!\times\!10^{-5}$ for $t \approx
2.6086$. Notice that $u(t,r)>-\pi$ for all times.}
\end{figure}
\begin{figure}
 \centering
\includegraphics[width=\textwidth]{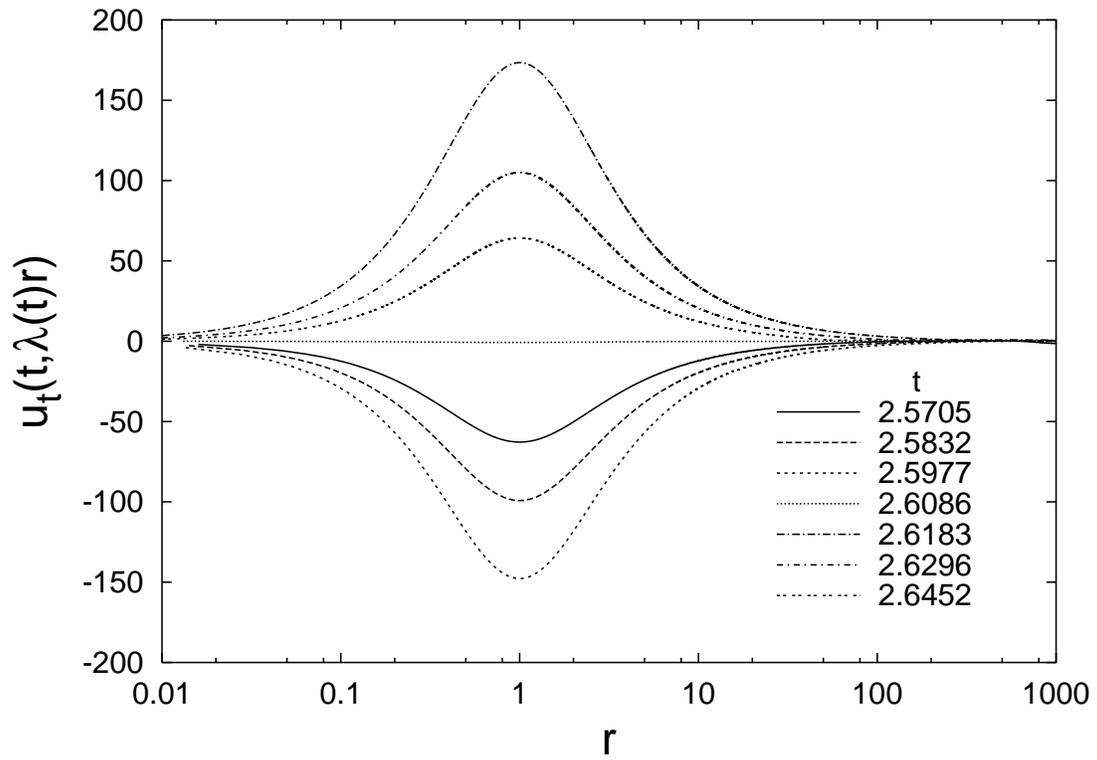}
\caption{The momenta for the same data as in Figure~3, rescaled by
the factors $\lambda(t)$. In agreement with the adiabatic
approximation (\ref{adiabatic}) the profiles have the shape of the
zero mode $v_0(r)$ with the amplitudes given by the logarithmic
derivative of the scale factor.}
\end{figure}
\begin{figure}
 \centering
\includegraphics[width=0.9\textwidth]{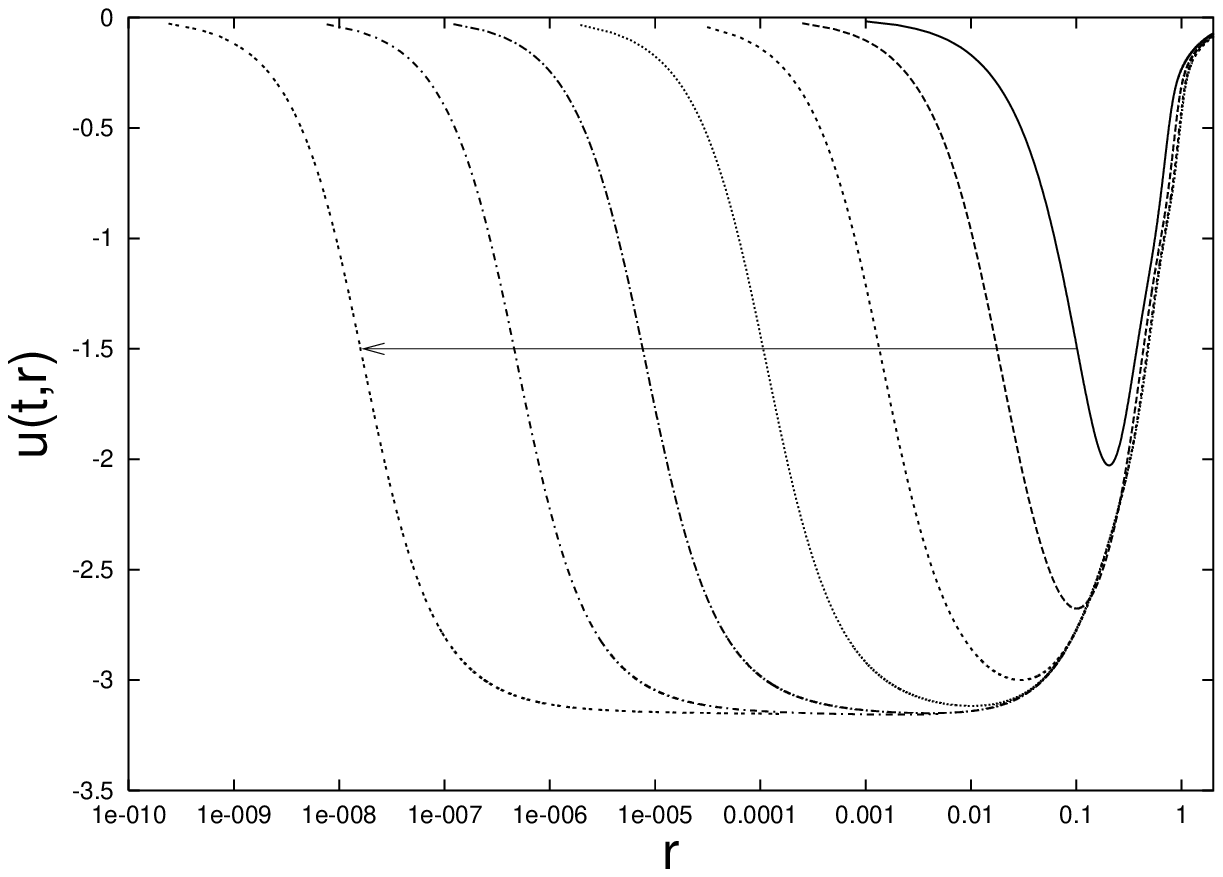}\\
\vspace{0.1in}%
\centering
\includegraphics[width=\textwidth]{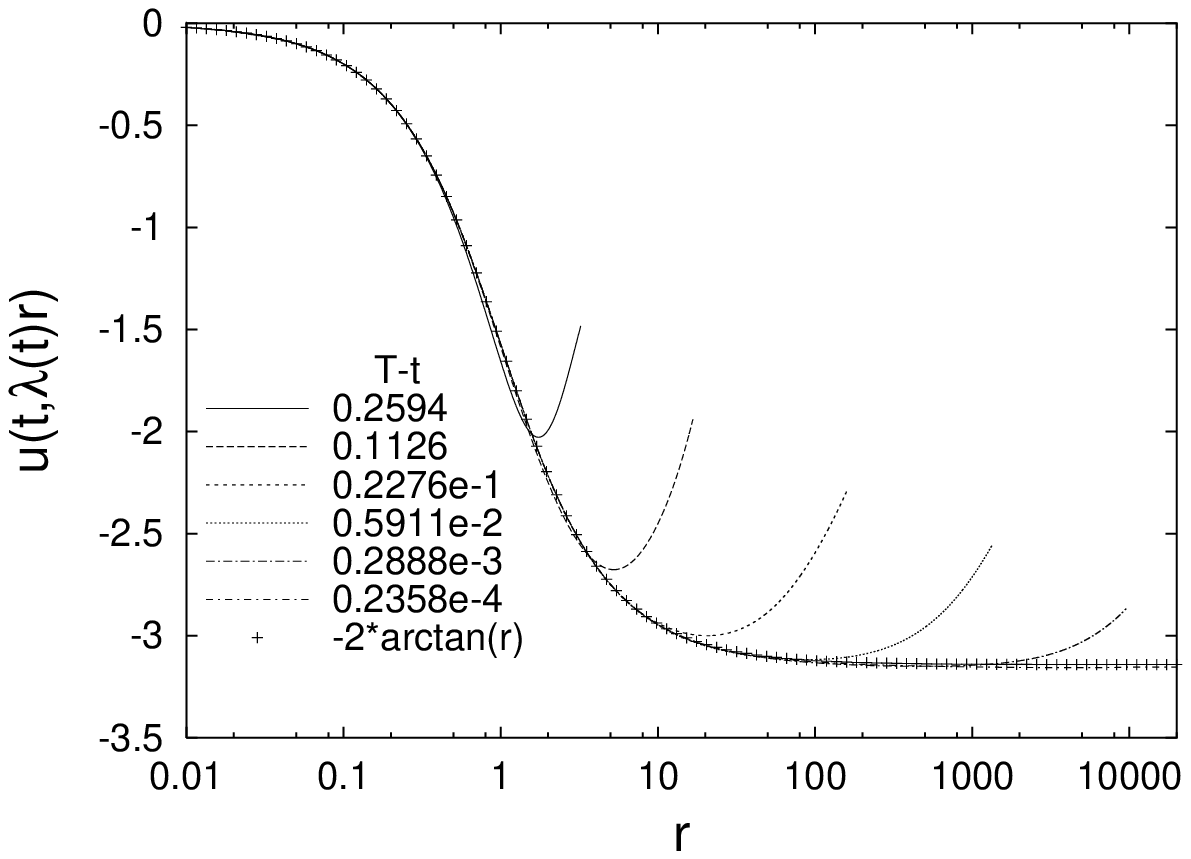}
\caption{Snapshots of the supercritical evolution for the
amplitude $A=1.072$. The upper plot shows the ingoing wave
shrinking indefinitely as $t$ approaches the blowup time $T
\approx 2.5593558$. In the lower plot the profiles from the upper
plot, rescaled by the factors $\lambda(t)$, are shown to collapse
to the static profile $-u_S(r)$. Notice that the fifth profile,
corresponding to $T-t=0.0002888$, overshoots $-\pi$; this seems to
be the necessary and sufficient condition for blowup, which is
reminiscent of  a similar phenomenon for the heat flow for
harmonic maps~\cite{pi}.}
\end{figure}
\begin{figure}
 \centering
\includegraphics[width=\textwidth]{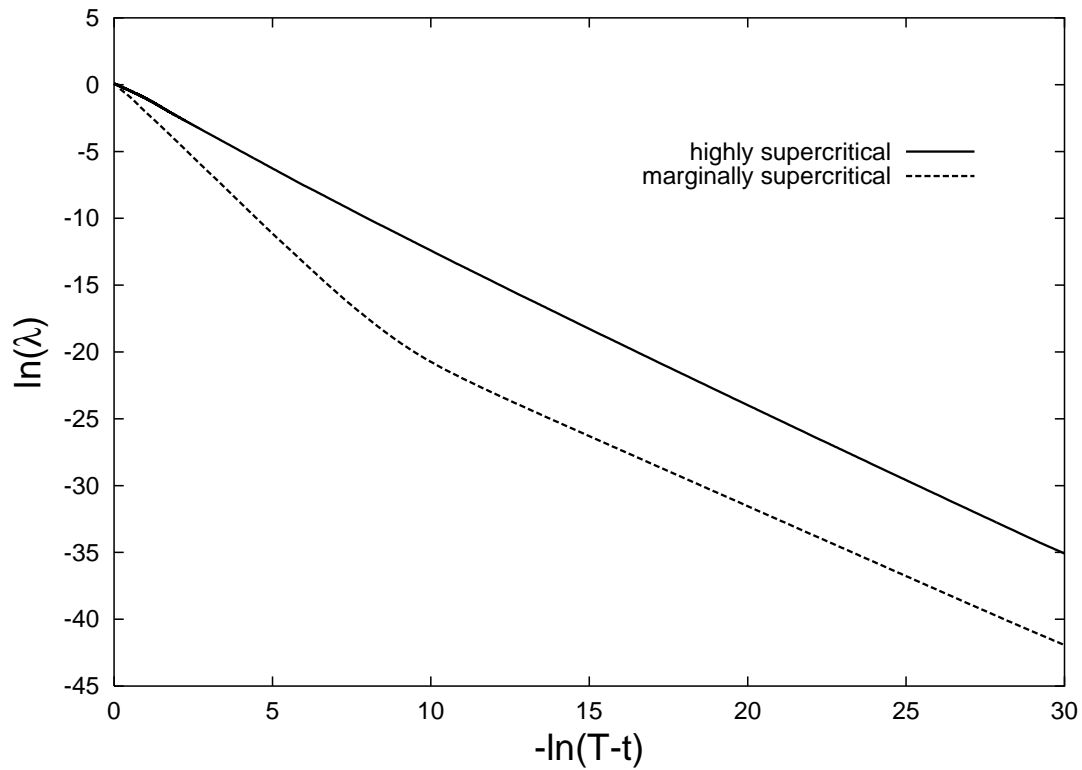}
\caption{The scale factor from two supercritical evolutions. In
both cases the asymptotic behavior of $\lambda(t)$ is well
approximated by the power law $\lambda \sim (T-t)^{\alpha}$ with
the exponent $\alpha \approx 1.1$. However, in the marginally
supercritical case the transient regime with $\alpha \approx 2.3$
in clearly seen before the true asymptotic state is reached.
As $A\rightarrow A^{\star}$, the crossover between the transient and the
asymptotic regimes occurs closer and closer to the blowup, which suggests
that solutions with exactly critical initial data blow up at the much slower
"transient-regime" rate.
}
\end{figure}
\end{document}